\documentclass[preprint,aps,prb,showpacs,amsmath]{revtex4}
\usepackage{latexsym}
\usepackage{amssymb}
\usepackage{graphicx}
\usepackage{bm}
\usepackage[english]{babel}
\usepackage[latin1]{inputenc}
\usepackage{amsmath}
\usepackage{amsfonts}

\newcommand{\pdag}{{\phantom{\dagger}}}
\newcommand{\bq}{\begin{equation}}
\newcommand{\eq}{\end{equation}}
\newcommand{\bn}{\begin{eqnarray}}
\newcommand{\en}{\end{eqnarray}}

\begin{document}

\title{Counting statistics of tunneling through a single molecule: effect of distortion and displacement of vibrational potential 
surface}

\author{Bing Dong}
\affiliation{Department of Physics, Shanghai Jiaotong University,
1954 Huashan Road, Shanghai 200030, China}

\author{H. Y. Fan}
\affiliation{Department of Physics, Shanghai Jiaotong University,
1954 Huashan Road, Shanghai 200030, China}

\author{X. L. Lei}
\affiliation{Department of Physics, Shanghai Jiaotong University,
1954 Huashan Road, Shanghai 200030, China}

\author{N. J. M. Horing}
\affiliation{Department of Physics and Engineering Physics, Stevens Institute of Technology, Hoboken, New Jersey 07030, USA}

\begin{abstract}

We analyze the effects of a distortion of the nuclear potential of a molecular quantum dot (QD), as well as a shift of its 
equilibrium position, on nonequilibrium-vibration-assisted tunneling through the QD with a single level ($\varepsilon_d$) coupled 
to the vibrational mode. For this purpose, we derive an explicit analytical expression for the Franck-Condon (FC) factor for a 
displaced-distorted oscillator surface of the molecule and establish rate equations in the joint electron-phonon representation 
to examine the current-voltage characteristics and zero-frequency shot noise, and skewness as well. 
Our numerical analyses shows that the distortion has two important effects. The first one is that it breaks the symmetry between 
the excitation spectra of the charge states, leading to asymmetric tunneling properties with respect to $\varepsilon_d>0$ and 
$\varepsilon_d<0$. Secondly, distortion (frequency change of the oscillator) significantly changes the voltage-activated cascaded 
transition mechanism, and consequently gives rise to a different nonequilibrium vibrational distribution from that of the case 
without distortion. Taken in conjunction with strongly modified FC factors due to distortion, this results in some new transport 
features: the appearance of strong NDC even for a single-level QD with symmetric tunnel couplings; a giant Fano factor even for a 
molecule with an extremely weak electron-phonon interaction; and enhanced skewness that can have a large negative value under 
certain conditions.       

\end{abstract}

\date{\today}


\maketitle

\section{Introduction}

Recently, the effect of unequilibrated quantized molecular vibrational modes on electronic tunneling has become an active issue, 
both experimentally\cite{jPark,hPark,Zhitenev,Yu,Pasupathy,LeRoy,Sapmaz} and 
theoretically.\cite{Bose,Alexandrov,McCarthy,Mitra,Koch,Koch1,Nowack,Koch2,Wegewijs,Zazunov,Haupt
,Dong,Shen,Merlo}
In particular, electronic transport measurements in suspended Carbon nanotubes have yielded $I$-$V$ curves demonstrating perfect 
signatures of phonon-mediated tunneling, e.g. stepwise structures with equal widths in voltage and gradual height reduction by 
the Franck-Condon (FC) factor.\cite{LeRoy,Sapmaz}  

To date, most theoretical works have taken into account only the displacement of the molecular oscillation arising from the 
entrance of an additional electron into the molecule. It is well known in molecular spectroscopy that upon an electronic 
transition, the potential surface involving the normal coordinates of the molecular oscillator generally undergoes a 
displacement, distortion, and rotation simultaneously. Accordingly, the molecular absorption and fluorescence line shapes depend 
on the square of the overlap integral between the initial wave function of the harmonic oscillator and the 
displaced-distorted-rotated one within the Born-Oppenheimer approximation, which is referred to as the FC factor in the 
literature.\cite{FC} Likewise, in the event of electron tunneling, an excess electron entering into a molecule from electrodes 
can induce a shift of the equilibrium position of the internal vibrational mode (displacement of the harmonic potential surface) 
and a simultaneous change of the vibrational frequency (distortion of the potential), both of which can modify the electronic 
tunneling rate and thus the nonlinear current-voltage characteristic of the molecule. (The rotation of the potential surface is 
ignored in this one-dimensional model.)   
J. Koch and F. von Oppen have studied this issue by calculating the overlap integral numerically in coordinate 
representation.\cite{Koch2} M.R. Wegewijs and K.C. Nowack have also analyzed this effect on electron tunneling by developing an 
explicit expression for the overlap integral in terms of Hermite polynomials.\cite{Wegewijs} However, this issue still requires 
further investigation. The main difficulty involves the derivation of an analytical expression for the vibrational overlap 
integral, which can be traced back to 1930\cite{Hutchisson} and has generated many studies in quantum 
chemistry.\cite{Herzberg1,Herzberg,Englman,Iachello} In this paper, we will employ Fan's method of integration within an ordered 
product of operators (IWOP) to derive an explicit analytical expression for the FC factors between two adiabatically 
displaced-distorted potential surfaces in terms of generalized Laguerre polynomials.\cite{Fan} The resulting exact FC factors 
will then be incorporated with the rate equation in a joint electron-phonon representation to analyze the effects of distortion 
on sequential tunneling through a single molecule. 

Furthermore, it is reasonable to expect that this distortion effect has an even stronger impact on current fluctuations, which 
has thus far not been studied despite much recent work investigating the unequilibrated phonon effect on current noise through a 
single molecule.\cite{Mitra,Koch,Koch1,Haupt,Shen} Recent developments in experimental measurement techniques for mesoscopic 
electron transport make it possible to detect the probability distribution of charge transmitted in a fixed time interval, namely 
the counting statistics of fluctuating electronic current.\cite{Bomze,Gustavsson,Fujisawa,Flindt,Levitov,Levitov1,Bednorz} Very 
recently, the full counting statistics (FCS) of tunneling current though a quantum dot (QD) were investigated in the Coulomb 
blockade regime by means of a rate equation.\cite{Bagrets,Emary,Braggio} Moreover, many studies have been devoted to the FCS of a 
single-level QD coupled to an oscillator with a harmonic potential.\cite{Flindt2} Therefore, another purpose of this paper is to 
study the counting statistics (up to the third moment) of electronic tunneling in a molecular QD based on number-resolved rate 
equations,\cite{Shen} focusing on the displaced-distorted effect (i.e. an oscillator with an anharmonic potential), the 
external-voltage-driven unequilibrated phonon effect, and finite phonon dissipation.

The rest of the paper is arranged as follows. In Sec.~II, we describe the model Hamiltonian and theoretical formulation of the 
rate equation. We briefly discuss how to incorporate the displacement-distortion effect into our previously derived rate 
equations in terms of a joint electron-phonon representation using a generic quantum Langevin equation approach with a Markovian 
approximation.
In this section, we also describe MacDonald's formulae for calculating the counting statistics, the first moment (i.e. {\em 
current}), and the zero-frequency second and third moments (i.e. the {\em shot noise} and the {\em skewness}, respectively).  
Then, we discuss in detail the counting statistics properties of a molecular QD for weak and strong distortion in Sec.~III. 
Finally, a brief summary is provided in Sec.~IV.

\section{Model and Theoretical Formulation}

\subsection{Model Hamiltonian}

We consider a simplified model of a molecular Hamiltonian constituted of two distinct parts, namely, electronic and nuclear parts 
within the Born-Oppenheimer approximation. The electronic part is assumed to be constituted of one spinless level, 
$\varepsilon_d$, with coupling to two electrodes [left (L) and right (R)] and also capacitively coupled to a gate electrode. 
Since an electron can tunnel into/out of this molecular QD from/into one of two electrodes in a transport process, there exist 
two possible electronic states: the molecular QD is either occupied by an electron ($|1\rangle$) or is unoccupied ($|0\rangle$). 
The nuclear part of the molecule is considered as a single harmonic oscillator. In this situation, there are two harmonic 
potentials corresponding to the two electronic states, which may differ from each other (we will term them the ground potential 
and the excited potential, respectively, in the following).
The full Hamiltonian of the molecule can be written as
\begin{subequations}
\bq
H = H_{leads}+H_{mol}+H_{T}, \label{hamiltonian}
\eq
with
\bn
H_{leads} &=& \sum_{\eta, {\bf k}} \varepsilon_{\eta {\bf k}} c_{\eta {\bf k}}^\dagger c_{\eta {\bf k}}^\pdag, \\
H_{mol} &=& \varepsilon_d |1\rangle \langle 1| + \sum_{N=0,1} |N\rangle \langle N|H_N, \\  
H_{N} &=& \frac{p^2}{2m_o} + \frac{m_o}{2} \omega_N^2 (x - d \delta_{N1} )^2 , \label{Hmol} \\
H_T &=& \sum_{\eta,{\bf k}} (V_{\eta} c_{\eta {\bf k}}^\dagger |0\rangle \langle 1| + {\rm H.c.}), \label{tunneling}
\en
\end{subequations}
where $c_{\eta{\bf k}}^\dagger$ ($c_{\eta{\bf k}}$) is the creation (annihilation) operator of an electron with momentum ${\bf 
k}$, and energy $\varepsilon_{\eta {\bf k}}$ in lead $\eta$ ($\eta=L,R$). $p$ ($x$) is the momentum (normal coordinate) of the 
molecular oscillator and $m_o$ is the mass. When there is no additional electron occupying the QD, the frequency of this 
oscillator is $\omega_0$ and its coordinate origin (the equilibrium position) is zero; but in the case of occupation, a 
distortion and a displacement occur, i.e., the frequency is changed to $\omega_1$ (which probably is different from $\omega_0$), 
and the equilibrium position has shifted through $d$, which is related to the coupling constant $\lambda=\omega_0 d \sqrt{m_o 
\omega_0 /2}$ for the electron-oscillator interaction. $V_\eta$ is the tunneling amplitude between the QD and electrode $\eta$. 
In this paper, we assume that $V_\eta$ is independent of the oscillator position, which is a good approximation for experimental 
realizations in most single molecule systems.\cite{jPark,hPark,Pasupathy,LeRoy,Sapmaz} We use units with $\hbar=k_B=e=m_o=1$ 
throughout the paper.

Bearing in mind that the rate equation is written in the joint electron-phonon representation, i.e., the direct product states of 
electron occupation number and phonon Fock state, it is convenient to rewrite the phononic part of the Hamiltonian, 
Eq.~(\ref{hamiltonian}), in quantized representation. We define phonon annihilation operators for the oscillator in two cases, 
$N=0$ and $1$, and phonon-bath, respectively, as:
\bq
a_N = \frac{1}{\sqrt{2}} \left [ \sqrt{\omega_N} (x - d \delta_{N1}) + \frac{ip}{\sqrt{\omega_N}} \right ], \label{aN}
\eq
In these terms, we have
\begin{subequations}
\bq
H_N = \omega_N a_N^\dagger a_N^\pdag ,
\eq
\end{subequations}
The definition of Eq.~(\ref{aN}) clearly shows that there are two different Fock spaces spanned by the harmonic oscillator basis 
sets, $|n\rangle_0$ and $|n\rangle_1$, corresponding to the ground and the excited harmonic potentials respectively,
\bq
a_N^\dagger a_N^\pdag |n\rangle_N = n |n\rangle_N.
\eq  
We will describe the electron-oscillator interaction in the tunneling Hamiltonian, $H_T$, using this representation. Consider a 
vacant initial electronic state of the molecule, $|0\rangle$, so its nuclear vibrational state is $|n\rangle_0$ of the ground 
potential, i.e., the initial joint electron-phonon state (JEPS) is $|0\rangle \otimes |n\rangle_0$ (for notational convenience, 
we use $0_n$ to denote this JEPS). If an electronic tunneling event occurs, an additional electron is injected into the QD from 
one of the electrodes, so that the electronic state changes to $|1\rangle$. In conjunction with this electronic transition, the 
nuclear vibrational state simultaneously changes to the Fock state $|n'\rangle_1$ of the excited potential due to external forces 
involved in the electronic transition in the molecule, leading to a final JEPS $|1\rangle \otimes |n'\rangle_1$ (denoted by 
$1_{n'}$). The total amplitude of the joint transition from the initial JEPS $0_n$ to the final JEPS $1_{n'}$ (which is called 
vibronic transition in the literature) is equal to the product of both the amplitudes of electronic tunneling to lead $\eta$, 
$V_\eta$, and the harmonic oscillator transition $|n\rangle_0 \rightarrow |n'\rangle_1$ with the FC factor $F_{n'n}$, which is 
determined by the overlap integral of the wavefunctions of the involved harmonic oscillator states in coordinate representation,
\bq
F_{nn'} = \int_{-\infty}^\infty dx \psi^*_{n}(x,\omega_0) \psi_{n'} (x-d,\omega_1)={}_0 \langle n|n'\rangle_1, 
\label{fc-definition}
\eq
where $\psi_{n}(x,\omega)$ denotes the eigenfunction of a harmonic oscillator with frequency $\omega$ in state $|n\rangle$. Using 
Fan's IWOP method, we can derive an explicit analytical expression for this FC overlap integral as (Appendix)\cite{Fan}
\begin{widetext}
\bn
F_{nn'} &=& (-1)^{n'} \exp \left [ -\frac{\gamma^2 \gamma'^2}{2(\gamma^2+\gamma'^2)}\right ] \left 
(\frac{2\beta}{1+\beta^2}\right )^{(n+n'+1)/2} \sum_{k=0}^{[n/2]} \sum_{k'=0}^{[n'/2]} \frac{\left 
(\frac{1-\beta^2}{4\beta}\right )^{k+k'}}{k!k'!} \frac{(-1)^{k'} (n! n'!)^{1/2}}{(n-2k)! (n'-2k')!} \cr
&& \times \left \{  
\begin{array}{cc}
(n'-2k')! (-1)^{n'} \left (\gamma' \sqrt{\frac{\beta}{1+\beta^2}}\right )^{n-2k-n'+2k'} L_{n'-2k'}^{n-2k-n'+2k'}\left 
(\frac{\beta \gamma \gamma'}{1+\beta^2}\right ), & \text{if}\,\, n-2k\geq n'-2k', \\
(n-2k)! (-1)^{n} \left (\gamma \sqrt{\frac{\beta}{1+\beta^2}}\right )^{n'-2k'-n+2k} L_{n-2k}^{n'-2k'-n+2k}\left (\frac{\beta 
\gamma \gamma'}{1+\beta^2}\right ), & \text{if}\,\, n-2k<n'-2k',
\end{array}
\right. \label{fc}
\en 
\end{widetext}
with $\beta=\sqrt{\omega_1/\omega_0}$, $\gamma=\sqrt{\omega_0} d=\sqrt{2}g$ ($g=\lambda/\omega_0$ is the dimensionless 
electron-vibration coupling constant), and $\gamma'=\sqrt{\omega_1} d=\beta \gamma$. $L_n^m(x)$ is the generalized Laguerre 
polynomial. Since the transport properties of the molecular QD depend crucially on the FC factors, it is essentially important in 
numerical calculations to compute these FC factors more accurately. The Eq.~(\ref{fc}) provides a more convenient and precise way 
to accomplish this than the pure summation expression in Ref.~\onlinecite{Iachello} and the direct numerical integration in 
Ref.~\onlinecite{Koch2}. Besides, unlike the earlier paper,\cite{Wegewijs} where the FC factor is expressed as a product of two 
Hermite polynomials, the Eq.~(\ref{fc}) here is a more compact and tractable formula for numerical computations.    
Finally, the tunneling Hamiltonian Eq.~(\ref{tunneling}) can be rewritten as
\bq
H_T = \sum_{\eta,{\bf k}} \sum_{n,n'} (V_{\eta} F_{nn'} c_{\eta {\bf k}}^\dagger |0\rangle |n\rangle_0 \, {}_1\langle n'| \langle 
1| + {\rm H.c.}). 
\eq

\subsection{Rate Equation}

Based on this representation of the Hamiltonian and following the same scheme we previously proposed,\cite{Shen} we have derived 
rate equations in terms of the JEPS representation of density matrix elements, $\rho_{00}^n=\langle |0\rangle |n\rangle_0 {}_0 
\langle n| \langle 0| \rangle$ and $\rho_{11}^n=\langle |1\rangle |n\rangle_1 {}_1 \langle n| \langle 1| \rangle$, to describe 
unequilibrated vibration-assisted sequential tunneling. It should be noted that coherences between state with different phonon 
number can be safely neglected owing to the big energy difference between the two JEPSs $0(1)_n$ and $0(1)_m$ if $n\neq m$ and a 
small level broadening due to tunneling.  
In order to employ MacDonald's formula for calculating counting statistics of the tunneling current, we write the rate equations 
in a number-resolved version describing the number of completed tunneling events,\cite{Shen}
\begin{subequations}
\bn
\dot \rho_{00}^{n(l)} &=& \sum_{m} \bigl [ \Gamma_{L,nm}^- \rho_{11}^{m(l)} 
 + \Gamma_{R,nm}^- \rho_{11}^{m(l-1)} \cr
&& - \Gamma_{nm}^+ \rho_{00}^{n(l)} \bigr ] - (\varpi_{0,n}^+ +\varpi_{0,n}^-) \rho_{00}^{n(l)} \cr
&& + \varpi_{0,n+1}^- \rho_{00}^{n+1(l)} + \varpi_{0,n-1}^+ \rho_{00}^{n-1(l)}, \label{ttqre:r00} \\
\dot \rho_{11}^{n(l)} &=& \sum_{m} \bigl [\Gamma_{L,mn}^+ \rho_{00}^{m(l)} + \Gamma_{R,mn}^+ \rho_{00}^{m(l+1)} \cr
&& - \Gamma_{mn}^- \rho_{11}^{n(l)} - (\varpi_{1,n}^+ +\varpi_{1,n}^-) \rho_{11}^{n(l)} \cr
&& + \varpi_{1,n+1}^- \rho_{11}^{n+1(l)} + \varpi_{1,n-1}^+ \rho_{11}^{n-1(l)}, \label{ttqre:r11} 
\en
\end{subequations}
in which the additional superscript $(l)$ denotes the total number of electrons transmitted through the QD that arrive at the 
right lead during a fixed time interval $t$. Obviously, we have the relation $\rho_{jj}^{n}(t)=\sum_{l} \rho_{jj}^{n(l)}(t)$ 
($j=0,1$). The electronic tunneling rates are defined as
\begin{subequations}
\bn
\Gamma_{nm}^+ &=& \sum_{\eta} \Gamma_{\eta ,nm}^+= \sum_{\eta} \Gamma_{\eta } \gamma_{nm} f_{\eta}(\varepsilon_d + 
m\omega_1-n\omega_0), \cr
&& \\
\Gamma_{nm}^- &=& \sum_{\eta} \Gamma_{\eta ,nm}^- \cr
&=& \sum_{\eta} \Gamma_{\eta } \gamma_{nm} [1-f_{\eta}(\varepsilon_d + m\omega_1-n\omega_0)],
\en
with $\Gamma_{\eta j}=2\pi \varrho_{\eta} |V_{\eta j}|^2$ denoting the tunneling strength between the QD and lead $\eta$ 
($\varrho_{\eta}$ is the density of states of lead $\eta$). $f_{\eta}(\epsilon)$ is the Fermi-distribution function of lead 
$\eta$ with temperature $T$, and the FC factors are
\bq
\gamma_{nm} = |F_{nm}|^2.
\eq
\end{subequations}
The dissipation rates of the vibrational number states are
\begin{subequations}
\bn
\varpi_{N,n}^+ &=& \varpi_p n_B(\omega_N) (n+1), \\
\varpi_{N,n}^- &=& \varpi_p (n_B(\omega_N)+1)n.
\en   
\end{subequations}
Here, we assume the same vibration-bath coupling constant, $\varpi_p$, for the vibrational modes.
In the case of a pure displacement of the vibrational potential ($\beta=1$), the FC factor becomes Eq.~(\ref{nodist}), and thus 
the rate equations Eqs.~(\ref{ttqre:r00}) and (\ref{ttqre:r11}) reduce to our previous results in Ref.~\onlinecite{Shen}.

\subsection{MacDonald's formula for couting statistics}

In principle, the full counting statistics of the tunneling current (all cumulants $\langle\langle q^l \rangle\rangle$ of the 
charge $q(t)$ transmitted through a QD during a sampling time $t$) can be calculated employing the number-resolved rate 
equations. However, the high moments are quite difficult to probe experimentally, due to the central limit theorem as well as 
their sensitivity to environmental influence.\cite{Levitov1} On the other hand, as a classical current fluctuation, a Gaussian 
distribution has all its cumulants higher than the second one equal to zero, i.e. $\langle\langle q^l \rangle\rangle\equiv 0$ for 
$l>2$; while a Poisson distribution has all its cumulants equal to its mean (the first cumulant). The simplest measurement of the 
non-Gaussianity of the distribution $q(t)$ is therefore the third current cumulant $\langle\langle q^3 \rangle\rangle$, which 
reflects the skewness of the distribution.
Hence, we focus on the first moment $\langle\langle q \rangle\rangle$ (the peak position of the distribution of 
transferred-electron-number), the second central moment $\langle\langle q^2 \rangle\rangle$ (the peak-width of the distribution), 
and the third central moment in this paper. 

According to the definition, the three cumulants can be calculated as
\begin{subequations}
\bn
\langle\langle q \rangle\rangle &=& \sum_{l} l P^{(l)}(t), \\
\langle\langle q^2 \rangle\rangle &=& \sum_{l} (l- \langle\langle q \rangle\rangle)^2 P^{(l)}(t) \cr
&=& \sum_{l} l^2 P^{(l)}(t) - \langle\langle q \rangle\rangle^2, \\ 
\langle\langle q^3 \rangle\rangle &=& \sum_{l} (l- \langle\langle q \rangle\rangle)^3 P^{(l)}(t) \cr
&=& \sum_{l} l^3 P^{(l)}(t) - 3\langle\langle q \rangle\rangle \langle\langle q^2 \rangle\rangle - \langle\langle q 
\rangle\rangle^3, 
\en
where
\bq
P^{(l)}(t) = \sum_{n} [\rho_{00}^{n(l)}(t)+ \rho_{11}^{n(l)}(t) ]
\eq
\end{subequations}
is the total probability of transferring $l$ electrons into the right lead by time $t$, which satisfies the normalization 
relation $\sum_l P^{(l)}(t)=1$. 

In the long time limit, these cumulants $\langle\langle q^l \rangle\rangle$ are proportional to the time, $\langle\langle q^l 
\rangle\rangle=\langle\langle {\cal I}^l \rangle\rangle t+c_l$, with the {\em current cumulants} $\langle\langle {\cal I}^l 
\rangle\rangle$ and constant terms $c_l$ ($c_1=0$).
From the definition, we have\cite{Shen,Dong3,MacDonald,Dong4}
\begin{subequations}
\bn
I &=& \frac{d}{dt} \langle\langle q \rangle\rangle {\Big |}_{t\rightarrow\infty}=\langle\langle \cal{I} \rangle\rangle, 
\label{Inr} \\
S_2 &=& \frac{d}{dt} \langle\langle q^2 \rangle\rangle {\Big |}_{t\rightarrow\infty} = \langle\langle {\cal I}^2 \rangle\rangle 
\cr
&=&\frac{d}{dt}\left [ \sum_{l} l^2 P^{(l)}(t) - (t I)^2 \right ] {\Big |}_{t\rightarrow\infty}, \label{sn}
\en
which are the average current and zero-frequency shot noise, respectively. Likewise, we can define the skewness as
\bn
S_3 &=& \frac{d}{dt} \langle\langle q^3 \rangle\rangle {\Big |}_{t\rightarrow\infty}= \langle\langle {\cal I}^3 \rangle\rangle  
\cr
&=& \frac{d}{dt}\left [ \sum_{l} l^3 P^{(l)}(t) - I^3 t^3 - 3 S_2 I t^2 -3Itc_2\right ] {\Big |}_{t\rightarrow\infty}, \label{sk}
\en
\end{subequations}

To evaluate $S_2$, $c_2$ and $S_3$, we define auxiliary functions $G_{jj}^{n}(t)$ and $H_{jj}^{n}(t)$ ($j=0,1$) as
\bn
G_{jj}^{n}(t) &=& \sum_{l} l \rho_{jj}^{n(l)}(t), \\
H_{jj}^{n}(t) &=& \sum_{l} l^2 \rho_{jj}^{n(l)}(t).
\en
Their equations of motion are readily derived employing the number-resolved QREs, Eqs~(\ref{ttqre:r00})--(\ref{ttqre:r11}), in 
matrix form: $\dot{\bm G}(t)={\cal M} {\bm G}(t) + {\cal G} {\bm \rho}(t)$ and $\dot{\bm H}(t)={\cal M} {\bm H}(t) + {\cal H} 
{\bm G}(t) + {\cal G}' {\bm \rho}(t)$ with ${\bm G}(t)=({\bm G}_{00}, {\bm G}_{11})^{T}$, ${\bm H}(t)=({\bm H}_{00}, {\bm 
H}_{11})^{T}$, and ${\bm \rho}(t)=({\bm \rho}_{00}, {\bm \rho}_{11})^{T}$ [here ${\bm G}_{jj}=(G_{jj}^{0}, G_{jj}^{1}, 
\cdots)^T$, ${\bm H}_{jj}=(H_{jj}^{0}, H_{jj}^{1}, \cdots)^T$, and ${\bm \rho}_{jj}=(\rho_{jj}^0, \rho_{jj}^1, \cdots)^T$]. 
${\cal M}$, ${\cal G}$, and ${\cal H}$, ${\cal G}'$ are easily obtained from Eqs.~(\ref{ttqre:r00})--(\ref{ttqre:r11}). 
Applying the Laplace transform to these equations yields
\bn
{\bm G}(s) &=& (s {\bm I}-{\cal M})^{-1} {\cal G} {\bm \rho}(s), \\
{\bm H}(s) &=& (s {\bm I}-{\cal M})^{-1} [{\cal H} {\bm G}(s) + {\cal G}'{\bm \rho}(s)], 
\en
with ${\bm \rho}(s)$ obtained by applying the Laplace transform to its constituent equations of motion using the initial 
condition ${\bm \rho}(0)={\bm \rho}_{st}$ (${\bm \rho}_{st}$ denotes the stationary solution of the rate equations). Due to the 
inherent long-time stability of the physical system under consideration, all real parts of nonzero poles of ${\bm \rho}(s)$, 
${\bm G}(s)$, and ${\bm H}(s)$ are negative definite. Consequently, divergent terms arising from the partial fraction expansions 
of ${\bm G}(s)$ and ${\bm H}(s)$ as $s\rightarrow 0$ entirely determine the large-$t$ behavior of the auxiliary functions, i.e. 
the zero-frequency shot noise and the skewness, Eqs.~(\ref{sn}) and (\ref{sk}).

\section{Results and discussion}

In this section, we present the results of our numerical investigation of nuclear distortion effects on tunneling current, $I$ 
[(in units of $(\Gamma_L+\Gamma_R)/\Gamma_L\Gamma_R$], and on low-frequency counting statistics, the normalized current noise 
(the Fano factor) $F_2=S_2/2I$ and the skewness $F_3=S_3/2I$ for the molecular QD. 

The vanishing of $\varpi_p=0$ denotes no dissipation of the vibration to the environment, corresponding to maximal uneqilibrated 
phonon effects in resonant tunneling; whereas increasing dissipation strength, $\varpi_p>0$, describes the action of the 
dissipative environment as it begins to relax the excited vibration towards an equilibrium state (the later occurs in the limit 
$\varpi_p=\infty$). Throughout the paper we set $\omega_0=1$ as the energy unit, $\mu_L=\mu_R=0$ at equilibrium condition, and we 
assume that the bias voltage $V$ is applied symmetrically, $\mu_L=-\mu_R=V/2$ with symmetrical tunnel couplings, 
$\Gamma_L=\Gamma_R=\Gamma$.

\subsection{Small shift of the equilibrium position}

To start, we consider cases involving a small displacement of the equilibrium position of the oscillator, $\lambda < 1$. 

\begin{figure}[htb]
\includegraphics[height=7.5cm,width=8.5cm]{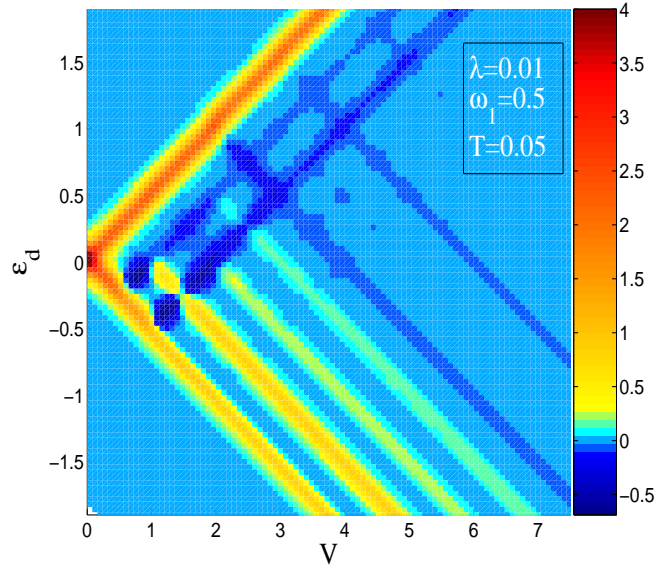}
\caption{(Color online). Differential conductance $dI/dV$ in the $V$-$\varepsilon_d$ plane for $\lambda=0.01$ and $\omega_1=0.5$ 
with $\Gamma_L=\Gamma_R$ and with no environmental dissipation. The temperature is $T=0.05$. The color bar gives the scale for 
the differential conductance.}
\label{fig3D1}
\end{figure}

The case with an extremely small shift $\lambda \ll 1$ ($\lambda=0.01$ in our calculations) corresponds to the situation in which 
charging the molecule induces only a distortion of the harmonic potential surface. It is already known that a pure displacement 
(no distortion, $\beta=1$) retains the symmetry of the $I$-$V$ curve with respect to the ground-state transition line 
($\varepsilon_d=0$), thus leading to a completely symmetrical $\varepsilon_d$-dependent differential conductance, 
$dI/dV$.\cite{Koch,Nowack,Shen} Moreover, there is no occurrence of negative differential conductance (NDC) for the single-level 
model with symmetric tunnel couplings.\cite{Nowack,Zazunov,Shen} However, a charging-induced distortion ($\beta\neq 1$) will 
indeed break the symmetry of the excitation spectra for $N=0$ (neutral state) and $N=1$ (anionic state), and thus also result in 
an asymmetric $dI/dV$-$\varepsilon_d$ curve due to the asymmetric FC factor $F_{nn'}$, as shown in Fig.~\ref{fig3D1}, which plots 
$dI/dV$ as a function of bias voltage $V$ and gate voltage $\varepsilon_d$ for $\omega_1=0.5$ at $T=0.05$. Surprisingly, this 
effect of pure distortion also gives rise to NDC even in the symmetric tunnel coupling model. These results are consistent with 
the earlier prediction of Ref.~\onlinecite{Wegewijs}.                          

In the following, we will provide detailed analyses of the effects of a pure distortion on the counting statistics of tunneling 
current and present an interpretation of the occurrence of NDC. We exhibit our calculated results without environmental 
dissipation, $\varpi_p=0$, for the $I$-$V$ characteristic and the bias-dependent Fano factor, $F_2$, for two systems, 
$\varepsilon_d=-0.5$ and $0.5$, with various different distortion ratios in Fig.~\ref{figif} at temperature $T=0.02$. It is 
evident that (1) strong NDC occurs for the case of $\varepsilon_d=-0.5$ and its pattern of NDC depends on the distortion ratio 
$\beta$; (2) by way of comparison, weak NDC is found for the case of $\varepsilon_d=0.5$; (3) the distortion causes asymmetric 
behavior of the gate-voltage-dependent higher orders of counting statistics, the shot noise $F_2$ and the skewness $F_3$ (see 
Fig.~6 below); (4) distortion induces an enhancement of shot noise and even a giant Fano factor $F_2$ up to $10^3$ for the case 
of $\varepsilon_d=-0.5$.

\subsubsection{The case with $\varepsilon_d<0$}\label{case1}

In an effort to understand these results, it is helpful to examine the bias voltage dependence of the joint electron-phonon 
occupation probabilities (JEPOPs) $\rho_{00}^n$ and $\rho_{11}^n$. We plot the calculated results in Fig.~\ref{figp1} for the 
system having $\varepsilon_d=-0.5$ with two different distortion ratios, $\omega_1=0.7$ [(a) and (b)] and $\omega_1=0.6$ [(c) and 
(d)]; and in Fig.~\ref{figp2} for the system having $\varepsilon_d=0.5$ with $\omega_1=0.7$ as well. Consider first the results 
for $\varepsilon_d=-0.5$ and $\omega_1=0.7$. It is clear from Fig.~\ref{figp1}(b) that the molecule is occupied by an electron 
(the state $1_0$) at the beginning and thus the current remains zero until $V\simeq 1.0 \omega_0$ (since no additional electron 
can enter into the molecule from the left lead). When the bias voltage increases to $1.0\omega_0$, the Fermi energy of the right 
lead is nearly equal to the energy of the JEPS $1_0$, leading to a de-population of the state $1_0$ and simultaneous populations 
of the states $0_0$ and $1_1$. That is to say, in this situation, only three JEPSs $0_0$, $1_0$, and $1_1$ are involved in 
tunneling events and all other JEPSs make no contribution to the tunneling, which is equivalent to a two-level QD with symmetric 
tunnel couplings ($x=\gamma_{00} \Gamma$ for the channel $1_0$ and $y=\gamma_{01} \Gamma$ for the channel $1_1$) under a large 
bias voltage at zero temperature.\cite{Dong4} Therefore, we have $\rho_{00}^0\simeq \rho_{11}^0\simeq \rho_{11}^1\simeq 1/3$, the 
current is $I=(x+y)\Gamma/3$, and the Fano factor is $F_2=[2(x^2+y^2)+xy]/(9xy)$. Due to the FC selection rule for vanishing 
shift ($x\simeq 0.98 \Gamma \gg y\sim 0$), we obtain a huge Fano factor, $F_2\simeq 2x/9y \sim 10^3$. 

When the bias voltage increases further to $1.4\omega_0$, the JEPSs $0_1$ and $1_3$ become occupied (surprisingly), albeit their 
conventional resonant values should be $V=2.0\omega_0$ and $2(3\omega_1+\varepsilon_d)=3.2\omega_0$, respectively. These unusual 
occupations can be understood qualitatively in terms of vibration-induced {\em cascaded} single-electron 
transitions:\cite{Nowack,Wegewijs,Shen} arbitrarily high vibrational excitations can in principle be accessed via {\em cascades} 
of single-electron tunneling processes, if these processes become energetically allowed by applying an appropriate bias voltage.                         
For instance, if the Fermi energy of the left lead is located at $0.7\omega_0$, the JEPS $1_1$ is populated and, moreover, the 
single-electron transition, $1_1\rightarrow 0_1$, is {\em permitted} because the bias voltage $V=1.4\omega_0$ provides sufficient 
energy to activate it, $\omega_0-(\varepsilon_d+\omega_1)=0.7\omega_0$. Concomitantly, the molecule is re-populated because the 
transition $0_1\rightarrow 1_3$ is {\em also} energetically accessible ($\varepsilon_d+3\omega_1-\omega_0=0.6\omega_0$), and it 
is allowed by the FC selection rule as well.
As a result, albeit the Fermi energy of the left lead is not aligned with the energy of the states $0_1$ and $1_3$, these states 
also become occupied [Fig.~\ref{figp1}(a,b)]. Another direct consequence of such cascade transitions is a decrease of the JEPOP 
$\rho_{11}^1$, and corresponding increase of the JEPOP $\rho_{11}^0$. Except for the exclusive relation between the states $0_0$ 
and $1_0$ at the beginning, they attract each other at sufficiently large bias voltage due to the strong bidirectional cascade 
transition ($1_0\leftrightarrow 0_0$) rate $\gamma_{00}\simeq 0.98$. Therefore, the molecule is further de-populated at this 
value of bias voltage and thus the current increases [Fig.~\ref{figif}(a)].   
Actually, it is evident from Figs.~\ref{figp1} and \ref{figp2} that $\rho_{11}^n=\rho_{00}^n$ is always satisfied for all $n$ at 
extremely large bias voltage since the permitted bidirectional transition $1_n\leftrightarrow 0_n$ makes the two JEPSs act as a 
whole.

Similarly, when bias voltage increases to $V\sim 1.8\omega_0$ ($\mu_L=0.9 \omega_0$), the JEPS $1_2$ becomes occupied, leading to 
simultaneous decreases of $\rho_{11}^0$ and $\rho_{00}^0$, as indicated by the double-arrow in Fig.~\ref{figp1}(a,b). As a 
result, the molecule has a higher probability of being occupied by an electron, causing a suppression of current at $V\sim 
1.8\omega_0$.

The situation is a little different for the case with $\omega_1=0.6$ as shown in Fig.~\ref{figp1}(c,d). For a small bias voltage 
$V \lesssim 1.0 \omega_0$, only three JEPSs $0_0$, $1_0$, and $1_1$ contribute to tunneling as discussed above in the case of 
$\omega_1=0.7$. We note the occurrence of the same values of current and huge Fano factor, $F_2\sim 10^3$ 
[Fig.~\ref{figif}(a,b)].
However, in contrast to the former case of $\omega_1=0.7\omega_0$, when the bias voltage is increased to $1.4\omega_0$, there are 
still only these three JEPSs involved in tunneling, since the bias voltage $1.4\omega_0$ can not provide sufficient energy to 
enable the transition $1_1\rightarrow 0_1$ [$\omega_0 - (\varepsilon_d+ \omega_1)= 0.9\omega_0$]. As a result, we find a further 
increase of $\rho_{11}^1$ at this value of bias voltage, as indicated by the double-arrow in Fig.~\ref{figp1}(c,d), and thus a 
corresponding decrease of $\rho_{00}^0$ leads to a NDC at the $I$-$V$ curve.  

\subsubsection{The  case with $\varepsilon_d>0$}

The case of positive $\varepsilon_d$ is quite different. In Fig.~\ref{figp2}, we plot the bias-voltage-dependent JEPOPs, 
$\rho_{00}^n$ and $\rho_{11}^n$ for the system with $\varepsilon_d=0.5$ and $\omega_1=0.7$. It is evident that (1) the QD begins 
to populate at $V\simeq 2|\varepsilon_d|\simeq 1.0 \omega_0$; (2) surprisingly, the states $0_1$ and $1_1$ are also nearly 
equally  populated at this point ($\rho_{00}^1\simeq \rho_{11}^1$ because of strong cascade transitions between the two states), 
albeit that the Fermi energy of the left lead is not matched with the energies of the states $0_1$ ($1.0 \omega_0$) and $1_1$ 
($1.2 \omega_0$); and there is not any cascade transition which can be activated to reach these two states. 

To explain these results analytically, we focus on the bias voltage region $1.0 \omega_0 <V<2.4 \omega_0$ and we consider an 
approximative model with only four accessible JEPSs, $0_{0,1}$ and $1_{0,1}$, at zero temperature. The simplified rate equations 
without environmental dissipation   read
\bn
\dot \rho_{00}^{0(l)} &=& -\Gamma_{00} \rho_{00}^{0(l)} 
 + \Gamma_{00} \rho_{11}^{0(l-1)}  + \Gamma_{01} \rho_{11}^{1(l)} +  \Gamma_{01} \rho_{11}^{1(l-1)}, \cr
\dot \rho_{00}^{1(l)} &=& -(\Gamma_{10}+ \Gamma_{11}) \rho_{00}^{1(l)} 
 + \Gamma_{10} \rho_{11}^{0(l-1)}  + \Gamma_{11} \rho_{11}^{1(l-1)} \cr
\dot \rho_{11}^{0(l)} &=& -(\Gamma_{00}+\Gamma_{10}) \rho_{11}^{0(l)} + \Gamma_{00} \rho_{00}^{0(l)} + \Gamma_{10} 
\rho_{00}^{1(l)} \cr
\dot \rho_{11}^{1(l)} &=& -(2\Gamma_{01}+\Gamma_{11}) \rho_{11}^{1(l)} + \Gamma_{11} \rho_{00}^{1(l)}, \label{rqsimple}
\en
with $\Gamma_{nm}=\Gamma \gamma_{nm}$. In this case, with $\lambda=0.01$ and $\omega_1=0.7$, we have $\Gamma_{00}=0.98\Gamma$ and 
$\Gamma_{11}=0.95\Gamma \gg \Gamma_{01}\approx 1\times 10^{-4}\Gamma$, and $\Gamma_{10}\approx 7\times 10^{-5}\Gamma$; and thus 
$\rho_{00}^0 \simeq \rho_{11}^0 \simeq \frac{1+2\Gamma_{01}/\Gamma_{10}}{4(1 + \Gamma_{01}/\Gamma_{10})} \approx 0.4$,
$\rho_{00}^1\simeq \rho_{11}^1 \simeq \frac{1}{4(1+\Gamma_{01}/\Gamma_{10})} \approx 0.1$; the current is $I \simeq 
\Gamma_{00}/2$, and the Fano factor is $F_2 \gtrsim 1$. Moreover, we find a little monotonic decreases of $\rho_{00}$ in the 
large bias voltage region, which is responsible for the weak NDC of the $I$-$V$ curve in Fig.~\ref{figif}(c). 

\subsubsection{Temperature and dissipation effects}

We have also examined the temperature and environmental dissipation dependences of the current and shot noise associated with a 
pure distortion effect. We plot the calculated results in Fig.~\ref{figif2} for the system with $\varepsilon_d=-0.5$, 
$\lambda=0.01$, and $\omega_1=0.7$. It should be noted that the NDC and huge Fano factor are quite fragile in regard to the 
environmental dissipation of the vibrational mode: the NDC nearly disappears and the Fano factor becomes $1/2$ (the typical value 
for a single-level QD with symmetric tunnel couplings) with a weak dissipation rate $\varpi_p=5\times 10^{-4} \omega_0$ 
[Fig.~\ref{figif2}(a,b)]; whereas the huge Fano factor is robust against increasing temperature (it remains $10^2$ up to a 
relatively high temperature over the range $T=0.2\omega_0\sim 0.5\omega_0$).    

\subsubsection{Skweness}

We analyze the {\em skewness or third moment} of counting statistics of tunneling event through a molecule due to unequilibrated 
vibration and pure distortion effect. The skewness characterizes the degree of asymmetry of a distribution around its mean, i.e., 
the distribution of transfered electron number around the average current $I$ here. A positive value of skewness signifies a 
electron transfer distribution with an asymmetric tail extending out towards more positive direction; while a negative value 
signifies a distribution whose tail extends out towards more negative direction.

The skewness can be easily obtained, using MacDonald's formula as described above, for a single level QD with no coupling to a 
vibrational mode at a sufficiently large bias voltage,
\bq
F_3=\frac{(\Gamma_L - \Gamma_R)^2 (\Gamma_L^2 + \Gamma_R^2) + 4\Gamma_L^2 \Gamma_R^2}{(\Gamma_L + \Gamma_R)^4},
\eq
which is consistent with the result of the zero-frequency limitation reported in Ref.~\onlinecite{Emary}. It reads $F_3^s=1/4$ 
for an symmetric system, $\Gamma_L=\Gamma_R$, and $F_3^{as}=1$ if $\Gamma_L\gg (\ll) \Gamma_R$, which shows that the transport 
behaves as a uncorrelated Poissonian event in the strongly asymmetric case. For a molecular QD with coupling to a vibrational 
mode, we observe from Fig.~\ref{figf31} (we set $\lambda=0.1$ in these calculations here) that, due to distortion effect, the 
skewness exhibits (1) asymmetric properties with respect to $\varepsilon_d$; (2) a large negative value at the bias voltage 
region corresponding to appearance of NDC for the system with $\varepsilon_d<0$; (3) but a positive value for the case with 
$\varepsilon_d>0$. For the case with $\varepsilon_d=-0.5$ and $\omega_1=0.7$, the system reduces to an equivalent two-level QD 
with symmetric couplings, $x$ and $y$ ($x\gg y$), respectively, as indicated in section \ref{case1}, at the bias voltage $V 
\lesssim 1.4 \omega_0$ and zero temperature. Therefore, we can derive an analytical expression for the skewness as
\bq
S_3= \frac{7x^2 y^2(x+y) -2 (x^5+y^5)}{27x^2y^2} \approx \frac{7x}{27} - \frac{2x}{27} \left (\frac{x}{y}\right )^2,  
\eq    
which gives a large negative value due to $x\gg y$. It is interesting to note that the emergence of a negative skewness in this 
system is completely stemming from the particular nonequilibrated population of the vibrational mode and the vibration-mediated 
tunneling rates at a certain bias voltage window, in comparison with that of a single level QD without coupling to a vibrational 
mode, in which the skewness is always positive at the whole bias voltage ranges. We believe that the appearance of large 
enhancement of the shot noise and the negative skewness is a distinct signature of the effects of nonequilibrated phonon and 
vibrational distortion on the transport dynamics of a molecular QD. 
Likewise, we can also derive the skewness for the system with $\varepsilon_d=0.5$ and $\omega_1=0.7$ based on 
Eq.~(\ref{rqsimple}), the full form of which is too cumbersome to give here. Instead, we assume $x=\Gamma_{00}\approx 
\Gamma_{11}$, $y=\Gamma_{01} \approx \Gamma_{10}$ and employ the fact, $x\gg y$, that leads to
\bq
S_3=\frac{5x^3}{16(x+y)^2}>0.
\eq 

We also examine the role of environmental dissipation on skewness in Fig.~\ref{figf31}(d). We observe that at equlibriated phonon 
situation $\varpi_p\rightarrow \infty$, the skewness arrives at the typical value $1/4$ of a symmetric system at large bias 
voltage region. Besides, the negative value of skewness for $\varepsilon_d \leq 0$ is robust against increasing temperature (not 
shown here).

\subsection{Large shift of the equilibrium position}

\subsubsection{Current and shot noise}

Here, we present numerical analyses of the effects on electronic tunneling of a large shift of the equilibrium position jointly 
with a finite distortion of the harmonic potential. It is already known that with no distortion, a strong shift of the potential 
can cause exponential suppression of the FC factors, and thus result in significant suppression of the vibration-modified 
electronic tunneling rates. This is to say that, a strong shift of the potential largely blocks electron tunneling in the low 
bias voltage region (FC blockade) and significantly enhances low-frequency shot noise.\cite{Mitra,Koch} The main effects of the 
distortion are the asymmetric dependence of differential conductance on $\varepsilon_d$ and appearance of weak NDC, as shown in 
the plot of the differential conductance as functions of $\varepsilon_d$ and bias voltage $V$ for a system with $\lambda=3.0$ and 
$\omega_1=0.8$ in Figs.~\ref{figf3D2} and \ref{figif3}(a).    

From Fig.~\ref{figif3}(a) and (b), it is evident that in absence of distortion ($\omega_1=1.0$), the large electron-phonon 
interaction strength induces significant current suppression and enhancement of shot noise due to the FC 
blockade.\cite{Mitra,Koch,Haupt,Shen} For instance, the effective tunneling rate between the states $0_0$ and $1_0$ is 
essentially reduced to $\Gamma_{00}=1.23\times 10^{-4} \Gamma$ owing to the FC factor for $\lambda=3.0$, which is responsible for 
electron trapping in the ground state of the QD, the state $1_0$, even at a resonant point of bias voltage, $V \gtrsim 
1.0\omega_0$ [as shown in Fig.~\ref{figp3}(a) and (b)]. It is also responsible for suppression of the current because $I_L 
\propto \Gamma_{00} \rho_{00}^0$ in the small bias voltage region. However, a distortion of the potential surface will increase 
this FC factor, $\Gamma_{00}$. For example, the value of $\Gamma_{00}=1.1\times 10^{-3} \Gamma$ for $\omega_1=0.6$ and 
$\lambda=3.0$ is one order of magnitude higher than that with no distortion (see Table \ref{tab:table2}). This is the reason that 
we observe a relatively weaker electron trapping effect in the ground state [Fig.~\ref{figp3}(c,d)] and an obviously enhanced 
current [Fig.~\ref{figif3}(a)] for a system with a distorted potential in comparison with those having no distortion.  

Moreover, the distortion effect induces an oscillatory-type structure of the occupation probabilities of the JEPSs as shown in 
Fig.~\ref{figp3}(c) and (d), in comparison to the stepwise probabilities in the case of no distortion [Fig.~\ref{figp3}(a) and 
(b)]. It is this oscillatory behavior of $\rho_{00}$ and $\rho_{11}$ that causes the appearance of NDC in $I$-$V$ curves as shown 
in Fig.~\ref{figf3D2} and Fig.~\ref{figif3}(a) and (c).
For the system with $\varepsilon_d=-0.5$, the molecule becomes empty at the bias voltage $V=1.0\omega_0$ for both situations. 
Without distortion, the sequential transition $1_0\leftrightarrow 0_0 \leftrightarrow 1_1 \leftrightarrow 0_1 \leftrightarrow 1_2 
\leftrightarrow 0_2 \cdots$ can be activated by the bias voltage $V \gtrsim 1.0\omega_0$ and arbitrarily high vibrational 
excitations can, in principle, be accessed via the cascades. Nevertheless, only the probabilities of the four states $0_{0(1)}$ 
and $0_{0(1)}$ are exhibited in Fig.~\ref{figp3}(a) and (b) for the bias voltage region $1.0\omega_0 \leq V < 3.0\omega_0$, while 
others are ignorable due to small transition rates. When the bias voltage increases to $3.0\omega_0$, some new cascaded 
transitions, e.g. $1_0 \leftrightarrow 0_1 \leftrightarrow 1_3 \leftrightarrow \cdots$, $0_0 \leftrightarrow 1_2 \leftrightarrow 
0_3 \leftrightarrow \cdots$, are available and these additional channels induce stepwise increases of $\rho_{00}$ and the 
current.  
If an additional electron induces a distortion of the oscillator of the molecule (we set $\omega_1=0.6$ here), the cascaded 
transition is quite different from that of the case of no distortion. When the bias voltage is about $1.0\omega_0 \sim 
1.4\omega_0$, the transition $1_1\leftrightarrow 0_1$ is not allowed because the bias voltage is not strong enough to trigger 
this tunneling event, $\omega_0-(\varepsilon_d+\omega_1)=0.9\omega_0$. Therefore, the infinite chain of the cascaded transition 
is broken, which means that in this bias voltage region, only three JEPSs $0_0$, $1_0$, and $1_1$ are involved in transport. In 
this situation, the tunneling can be described by a simplified symmetric two-level model, as indicated in Sec.~\ref{case1}, with 
$x=\gamma_{00}=1.1\times 10^{-3}$ and $y=\gamma_{01}=9.5\times 10^{-3}$ (see table \ref{tab:table2}), which reveals a small 
current, $I\sim (x+y) \Gamma/3$, an enhanced shot noise, $F_2\sim 10$, and a very small negative skewness [see 
Fig.~\ref{figf34}(a) below].                   

However, when the bias voltage increases above $1.4\omega_0$, another transition, $0_0 \leftrightarrow 1_2$, is activated with a 
stronger transition strength, $\gamma_{02}=3.8\times 10^{-2}$, leading to a decrease of $\rho_{00}^0$and to population of the 
state $1_2$, which also results in the opening of the transition, $1_2 \leftrightarrow 0_1$, with a stronger transition strength, 
$\gamma_{12}=9.2\times 10^{-2}$ [because the energy change of this transition is quite small, $\omega_0- (\varepsilon_d + 
2\omega_1)=0.3\omega_0$], giving rise to population of the state $0_1$, as shown in Fig.~\ref{figp3}(c) and (d). This result is 
also different from that of the pure distortion case considered in Sec.~\ref{case1}, where the transition, $1_2\leftrightarrow 
0_1$, is not allowed due to the FC selection rule and the state $0_1$ is not populated until the bias voltage $V=1.8\omega_0$ is 
reached [Fig.~\ref{figp1}(c) and (d)]. If the bias voltage is increased to $1.8\omega_0$, the transition, $1_1 \leftrightarrow 
0_1$ is opened, which causes (1) decrease of $\rho_{11}^1$; (2) further increase of $\rho_{00}^1$; and (3) a feedback effect: 
increases of $\rho_{00}^0$ and $\rho_{11}^0$ due to the cascaded back-transition channel, $0_1\rightarrow 1_2 \rightarrow 0_0 
\rightarrow 1_0$. At the bias voltage $V=2.6\omega_0$, two more transitions, $0_0\leftrightarrow 1_3 \leftrightarrow 0_2$ and 
$0_0\leftrightarrow 1_2 \leftrightarrow 0_2$, are activated, resulting in a decrease of $\rho_{00}^0$ again. At $V=3.0\omega_0$, 
the state $1_0$ can directly transit to the state $0_1$: This new transition induces a strong decrease of $\rho_{11}^0$ and its 
strong feedback effect makes $\rho_{00}^0$, $\rho_{00}^1$, and $\rho_{11}^1$ all increase. To sum up, the newly opened cascaded 
transitions and their strong feedback effects are responsible for the oscillatory behavior of $\rho_{00}$ and $\rho_{11}$. 

Environmental dissipation will gradually destroy the NDC and the super-Poissonian shot noise, as shown in Fig.~\ref{figif3}(c) 
and (d). Unlike the case of pure distortion [Fig.~\ref{figif2}(a)], the current is obviously reduced by environmental 
dissipation.   
 
\subsubsection{Skweness}

We have also examined the skewness of the vibrational-mediated electron tunneling in the case of a large shift of the equilibrium 
position of the oscillator, both with and without distortion. Figure \ref{figf33} shows the skewness $F_3$, in the case of no 
distortion, as a function of bias voltage for systems with $\varepsilon_d=0$ and $\lambda=1.0$ (a) and $\lambda=3.0$ (b). Unlike 
the case of pure distortion, there is no occurrence of negative values of $F_3$. It is clear that in the equilibrated phonon 
situation (infinite environmental dissipation rate, $\varpi_p\rightarrow \infty$), the skewness has the typical value, 
$F_3^s=1/4$, for the symmetric single-level QD. For a moderate shift, $\lambda=1.0$, the skewness is slightly larger than $F_3^s$ 
in the small bias voltage region, but it is smaller than $F_3^s$ in the large bias voltage region. More interestingly, a finite 
dissipation rate obviously lowers the skewness below $F_3^s$ even in the small bias voltage region. Moreover, we find a 
significant enhancement (up to $10^2$) of the skewness in the case of a large shift, $\lambda=3.0$, which can also be ascribed to 
the FC blockade effect.         

In the presence of distortion (Fig.~\ref{figf34}), the skewness exhibits asymmetry with respect to $\varepsilon_d$ and obviously 
gives rise to negative values in a moderate bias voltage region. For a system with $\varepsilon_d=-0.5$ [Fig.~\ref{figf34}(a)], 
distortion even results in a small negative skewness in a small bias voltage, as mentioned above. Environmental dissipation 
destroys these features of skewness (not shown here).

\section{Conclusions}   

In summary, we have analyzed, in great detail, the effect of distortion of the harmonic potential surface of a molecular 
oscillator (due to additional electron occupation) on unequilibrated vibration-assisted tunneling through a molecular QD in the 
sequential tunneling regime. This effect is modelled by a change of the vibrational frequency (phonon energy) in addition to a 
shift of the equilibrium position of the potential during electronic tunneling. To take this effect into account, we developed an 
explicit analytical expression for the vibrational overlap integral involving the harmonic wavefunction jointly with its 
shifted-distorted counterpart, which is known as the FC factor in literature. 
It is this factor that significantly modifies the tunneling rates during electron hopping between the electrodes and the 
molecule, and consequently it strongly influences the electronic tunneling properties of such molecular devices. In this paper, 
we employed Fan's IWOP technique to accomplish this task.

Using these derived analytical expressions for the FC factors, we have established generic rate equations in terms of the JEPS 
representation to describe vibration-mediated tunneling, which facilitated examination of the roles of both shifting and 
distortion of the oscillator, as well as the roles of the unequilibrated phonon, and its dissipation to environment. 
Employing MacDonald's formula, we have calculated the current and its counting statistics, i.e., the low-frequency shot noise and 
skewness as well. 
         
Our analyses show that distortion has two main effects.\cite{Wegewijs} The first one is that, due to distortion-induced symmetry 
breaking between the excitation spectra of the two charge states of the molecule, i.e., the neutral state with $N=0$ and the 
anionic state with $N=1$, the FC factors are asymmetric ($\gamma_{nm}\neq \gamma_{mn}$ if $n\neq m$) and the molecular QD 
exhibits asymmetric tunneling properties with respect to $\varepsilon_d>0$ and $\varepsilon_d<0$. The second one 
is that, because the change of oscillator frequency can significantly change the cascaded transition channels activated by an 
external bias voltage, it gives rise to different nonequilibrium vibrational distributions, and thus results in some new 
tunneling properties which  are absent from the molecular QD without distortion.       

In particular, for the system of pure distortion (we suppose an extremely small shift of equilibrium position in numerical 
calculations), we have found that the molecular QD exhibits a strong NDC even at the condition of symmetric tunnel couplings, an 
enhanced shot noise with a huge Fano factor, and a large negative skewness in an anionic state; while it only has a weak NDC, a 
nearly Poissonian shot noise, and a large positive skewness for a neutral molecule. Two simplified models (and thus simplified 
rate equations) have been given to interpret qualitatively the reasons of different transport properties. In the presence of a 
large shift of the potential minima, the distortion also causes appearance of weak NDC and an enhanced shot noise, which can be 
ascribed to oscillatory behavior of the probabilities of the JEPSs due to the modified cascaded transition and strong feedback 
effects. The strongly enhanced skewness may be also negative at certain bias voltage.

\begin{acknowledgments} 

We gratefully thank Mr. Li-Yun Hu for his derivation of the FC factor, Eq.~(\ref{fc}). This work was supported by Projects of the 
National Science Foundation of China, Specialized Research Fund for the Doctoral Program of Higher Education (SRFDP) of China, 
and the Program for New Century Excellent Talents in University (NCET).

\end{acknowledgments}

\appendix*

\section{Derivation of Eq.~(\ref{fc})}

For reference purposes, we briefly review the derivation of Eq.~(\ref{fc}) following the Fan's IWOP method.\cite{Fan}

In a FC transition, e.g. electronic tunneling through a harmonic oscillator in the present paper, it is usually supposed that the 
initial state of the molecule (corresponding to the empty electronic state of the neutral molecule) is a standard harmonic 
oscillator, $H_0$ [Eq.~(\ref{Hmol})]; whereas after an electronic transition occurs (the anionic molecule), the harmonic 
vibrational potential of the molecule, $H_1$, is distorted and displaced from that of the neutral one. Considering Eq~(\ref{aN}), 
we have two different Hilbert spaces, $|n\rangle_0$ and $|n\rangle_1$, and the FC transition is described by 
Eq.~(\ref{fc-definition}).

As suggested by Fan,\cite{Fan} Eq.~(\ref{fc-definition}) can also be rewritten as
\bn
F_{nn'} &=& \int_{-\infty}^{+\infty} dx \,\,{}_0 \langle n | x \rangle_{\omega_0} \cdot {}_{\omega_1} \langle x-d | n' \rangle_0 
\cr
&=& {}_0 \langle n | \int_{-\infty}^{+\infty} dx \,| x\rangle_{\omega_0} \cdot {}_{\omega_1} \langle x-d | n' \rangle_{0},
\en   
in which $|x\rangle_{\omega_0}$ and $|x-d\rangle_{\omega_1}$ are basis vectors in coordinate representation. Their expressions in 
Fock representation are
\bq
|x\rangle_{\omega_0} = \left (\frac{\omega_0}{\pi}\right )^{1/4} \exp \left ( -\frac{\omega_0}{2} x^2 + \sqrt{2\omega_0} x 
a_0^\dagger - \frac{1}{2} a_0^{\dagger \,2} \right ) |0\rangle_{0},
\eq
\bq 
|x\rangle_{\omega_1} = \left (\frac{\omega_1}{\pi}\right )^{1/4} \exp \left ( -\frac{\omega_1}{2} x^2 + \sqrt{2\omega_1} x 
a_0^\dagger - \frac{1}{2} a_0^{\dagger \,2} \right ) |0\rangle_{0}, \label{x1}
\eq
with $|0\rangle_{0}$ as the vacuum state of $a_0$ ($a_0 |0\rangle_{0} =0$). It is convenient to define the dyadic (ket-bra) 
operator, $S$, as
\bn
S &=& \int_{-\infty}^{+\infty} dx \,\, | x-d \rangle_{\omega_1} \cdot {}_{\omega_0} \langle x | = \left ( \frac{\omega_0 
\omega_1}{\pi} \right )^{1/4} \int_{-\infty}^{+\infty} dx \cr 
&& \times \exp \left [ -\frac{\omega_1}{2} (x-d)^2 + \sqrt{2\omega_1} (x-d) a_0^\dagger - \frac{a_0^{\dagger \, 2}}{2} \right ] 
\cr
&& \times  |0\rangle_0 \, {}_0\langle 0| \exp \left [ -\frac{\omega_0}{2} x^2 + \sqrt{2\omega_0} x a_0 - \frac{a_0^{2}}{2} \right 
].
\en
Employing the operator formulae,
\bn
|0\rangle_0 \, {}_0\langle 0| &=& : \exp (-a_0^\dagger a_0^\pdag):,\\
\exp(c a_0^\dagger a_0^\pdag) &=& : \exp [(e^c-1) a_0^\dagger a_0^\pdag]:,
\en 
($:\cdots:$ denotes normal ordering of the operators within the colons and $c$ is a constant), we find
\bn
S &=& \left (\frac{\omega_0 \omega_1}{\pi^2}\right )^{1/4} \int_{-\infty}^{+\infty} dx \, : \exp \left [ -\frac{x^2}{2} 
(\omega_0+\omega_1) \right. \cr
&& + \sqrt{2} x \left ( \sqrt{\omega_1} a_0^\dagger + \sqrt{\omega_0} + \frac{\omega_1 d}{\sqrt{2}} \right ) \cr
&& \left. - \frac{\omega_1}{2} d^2 -\sqrt{2\omega_1} d a_0^\dagger - \frac{1}{2} (a_0+a_0^{\dagger})^2 \right ]: \cr
&=& e^{- \frac{\gamma^2 \gamma'^2}{2(\gamma^2+\gamma'^2)}} \exp \left [ -\frac{\tanh \chi}{2} a_0^{\dagger \,2} - \frac{{\rm 
sech}\, \chi}{\sqrt{2}} \gamma a_0^\dagger \right ] \cr
&& \times \exp \left [ \left ( a_0^\dagger a_0^\pdag + \frac{1}{2}\right ) \ln {\rm sech} \, \chi \right ] \cr
&& \times \exp \left ( \frac{a_0^2}{2} \tanh \chi + \frac{{\rm sech} \, \chi}{\sqrt{2}} \gamma' a_0 \right ),
\en
with $\chi=\ln \beta$.
It is easy to verify that 1) $S$ is a unitary operator; 2) $S|x\rangle_{\omega_0} = |x-d\rangle_{\omega_1}$, which means that $S$ 
is an operator that transforms a state of the oscillator into a distorted-displaced state (FC transition); and 3) $S^{\dagger} 
a_0 S = a_0 \cosh \chi - a_0^\dagger \sinh \chi - \gamma'/\sqrt{2}=a_1$.   

Using the operator $S$, the FC transition can be represented as follows,
\bq
F_{nn'}= {}_0\langle n | S^\dagger |n'\rangle_0,
\eq
which can be evaluated in the coherent state representation. 
Noting the relation between number states and the coherent state
\bq
|n\rangle_{0} = \frac{a_0^{\dagger \, n}}{\sqrt{n!}} |0\rangle_{0} =\frac{1}{\sqrt{n!}} \left ( \frac{d}{d\alpha} \right )^n 
|\alpha\rangle {\Big |}_{\alpha=0},
\eq
we have
\bq
F_{nn'} = (n! n'!)^{-1/2} \left ( \frac{d}{d\alpha^*} \right )^n \left ( \frac{d}{d\tau} \right )^{n'} \langle \alpha |S^\dagger| 
\tau \rangle {\Big |}_{\alpha^*=\tau=0}. \label{fnn'} 
\eq

Considering two two coherent states $\alpha$ and $\tau$, we have
\bn
&& a_0 |\alpha\rangle = \alpha |\alpha\rangle, \,\,\langle \alpha|a_0^\dagger = \alpha^* \langle \alpha|, \cr
&& |\alpha\rangle = e^{\alpha a_0^\dagger} |0\rangle_{0},\,\,\langle \alpha|\tau \rangle=e^{\alpha^* \tau}, \cr
&& \langle \alpha |:f(a_0, a_0^\dagger) :|\tau \rangle =f(\tau,\alpha^*) e^{\alpha^* \tau}, 
\en
and, consequently, we can derive
\bn
\langle \alpha |S^\dagger |\tau\rangle &=& A \exp \left [ \frac{\tanh \chi}{2} (\alpha^{*\,2} - \tau^2) + \alpha^* \tau {\rm 
sech}\, \chi \right. \cr
&& \left. + \frac{{\rm sech} \, \chi}{\sqrt{2}} (\gamma' \alpha^* - \gamma \tau) \right ], \label{sat}
\en
where
\bq
A = {}_0\langle 0|S^\dagger |0\rangle_{0}= {\rm sech}^{1/2} \chi \exp \left [-\frac{\gamma^2 
\gamma^{'2}}{2(\gamma^2+\gamma^{'2})} \right ].
\eq
Substituting Eq.~(\ref{sat}) into Eq.~(\ref{fnn'}) and using the definition of the double-variable Hermite 
polynomial,\cite{Erdelyi} 
\bn
H_{m,n}(\xi,\varsigma) &=& \frac{\partial^{m+n}}{\partial t^m \partial t'^n} \exp [-tt' + \xi t+ \varsigma t'] |_{t=t'=0} \cr
&=& \sum_{l=0}^{min(m,n)} \frac{m!n!(-1)^l}{l! (m-l)! (n-l)!} \xi^{m-l} \varsigma^{n-l},\cr
&& \label{Hmn}
\en
and the relation
\bq
\frac{\partial^{l+k}}{\partial \xi^l \partial \varsigma^k} H_{m,n}(\xi,\varsigma)= \frac{m!n!}{(m-l)! (n-k)!} 
H_{m-l,n-k}(\xi,\varsigma), 
\eq
we find
\bn
F_{nn'} &=& (-1)^{n'} e^{-\frac{\gamma^2 \gamma'^2}{2(\gamma^2+\gamma'^2)}} \left (\frac{2\beta}{1+\beta^2}\right )^{(n+n'+1)/2} 
\cr
&& \times \sum_{k=0}^{[n/2]} \sum_{k'=0}^{[n'/2]} \frac{\left (\frac{1-\beta^2}{4\beta}\right )^{k+k'}}{k!k'!} 
 \frac{(-1)^{k'} (n! n'!)^{1/2}}{(n-2k)! (n'-2k')!} \cr
&& \times H_{n-2k,n'-2k'} \left (\gamma' \sqrt{\frac{\beta}{1+\beta^2}}, \gamma \sqrt{\frac{\beta}{1+\beta^2}} \right ). \cr
&&
\en 
Finally, employing the relation between Hermite polynomials and Laguerre polynomials,
\bq
H_{m,n}(\xi,\varsigma)= \left \{ 
\begin{array}{cc}
(-1)^n n! \xi^{m-n} L_n^{m-n}(\xi\varsigma), & \,\, m\geq n, \\
(-1)^m m! \varsigma^{n-m} L_m^{n-m}(\xi\varsigma), & \,\, m< n,
\end{array}
\right.
\eq
we obtain Eq.~(\ref{fc}).

In particular, for $\beta=1$ (no distortion), it is readily verified that Eq.~(\ref{fc}) reduces to the well-known expression 
[see Ref.~\onlinecite{Shen}, Eq.~(9)]
\bq
F_{nn'} = \left \{
\begin{array}{cc}
e^{-g^2/2} g^{n'-n} \sqrt{\frac{n!}{n'!}} L_n^{n'-n} (g^2), & n< n', \\
e^{-g^2/2} (-g)^{n-n'} \sqrt{\frac{n'!}{n!}} L_{n'}^{n-n'} (g^2), & n\geq n';
\end{array}
\right. \label{nodist}
\eq
whereas for $d=0$ (no shift of equilibrium position), this case is relevant to the theory of activity of non-totally symmetric 
vibrations in the electronic spectra of polyatomic molecules.\cite{Herzberg} We easily obtain the Herzberg-Teller selection rule, 
$F_{nn'}=0$ if $n-n' \neq {\rm even}$.\cite{Herzberg1} Moreover, we find
\bn
F_{n0} &=& \frac{(-i)^{n}}{(n!)^{1/2}} \exp \left [ -\frac{\gamma^2 \gamma'^2}{2(\gamma^2+\gamma'^2)}\right ] \left 
(\frac{2\gamma\gamma'}{\gamma^2 + \gamma'^{2}}\right )^{1/2} \cr
&& \times \left (\frac{\gamma^2 - \gamma'^2}{2(\gamma^2 + \gamma'^2)}\right )^{n/2} H_{n}\left ( \frac{i \beta 
\gamma'}{\sqrt{1-\beta^4}}\right ),    
\en
and 
\bn
F_{0n'} &=& \frac{(-1)^{n'}}{(n'!)^{1/2}} \exp \left [ -\frac{\gamma^2 \gamma'^2}{2(\gamma^2+\gamma'^2)}\right ] \left 
(\frac{2\gamma\gamma'}{\gamma^2 + \gamma'^{2}}\right )^{1/2} \cr
&& \times \left (\frac{\gamma^2 - \gamma'^2}{2(\gamma^2 + \gamma'^2)}\right )^{n'/2} H_{n'}\left ( \frac{\beta 
\gamma}{\sqrt{1-\beta^4}}\right ),    
\en
in which $H_{n}(x)$ is the single variable Hermite polynomial. These special results are identical to those of Englman's 
derivation.\cite{Englman}

\newpage

\begin{table}
\caption{\label{tab:table2} Several FC factors $\gamma_{nm}$ ($n,m=0\sim 3$) with $\lambda=3.0$ and $\omega_1=0.6$ ($\gamma_{nm} 
\neq \gamma_{mn}$).}
\begin{ruledtabular}
\begin{tabular}{c|cccc}
& $m=0$ & 1 & 2 & 3 \\
\hline
$n=0$ & $1.1\times 10^{-3}$ & $9.5\times 10^{-3}$ & $3.8\times 10^{-2}$ & $9.4\times 10^{-2}$ \\
$1$ & $5.7\times 10^{-3}$ & $3.5\times 10^{-2}$ & $9.2\times 10^{-2}$ & $0.132$ \\
$2$ & $1.6\times 10^{-2}$ & $6.95\times 10^{-2}$ & $0.114$ & $7.38\times 10^{-2}$ \\
$3$ & $3.0\times 10^{-2}$ & $9.6\times 10^{-2}$ & $8.5\times 10^{-2}$ & $1.0\times 10^{-2}$ \\
\end{tabular}
\end{ruledtabular}
\end{table}

\vspace*{2cm}

\newpage

\centerline{\Large Figure Caption}

\vspace{2cm}

\noindent FIG.1: Differential conductance $dI/dV$ in the $V$-$\varepsilon_d$ plane for $\lambda=0.01$ and $\omega_1=0.5$ with 
$\Gamma_L=\Gamma_R$ and with no environmental dissipation. The temperature is $T=0.05$. The color bar gives the scale for the 
differential conductance.

\vspace{1cm} 

\noindent FIG.2: (Color online). Distortion effect on the bias voltage dependent current, $I$ (a,c) and Fano factor, $F_2=S_2/2I$ 
(b,d) for the case of an extremely weak shift of the equilibrium position of the vibrational potential, $\lambda=0.01$ with 
$\varepsilon_d=-0.5$ (a,b) and $\varepsilon_d=0.5$ (c,d), $\Gamma_L=\Gamma_R$ and with no environmental dissipation. The 
temperature is $T=0.02$.

\vspace{1cm} 

\begin{figure}[htb]
\includegraphics[height=7.5cm,width=8.5cm]{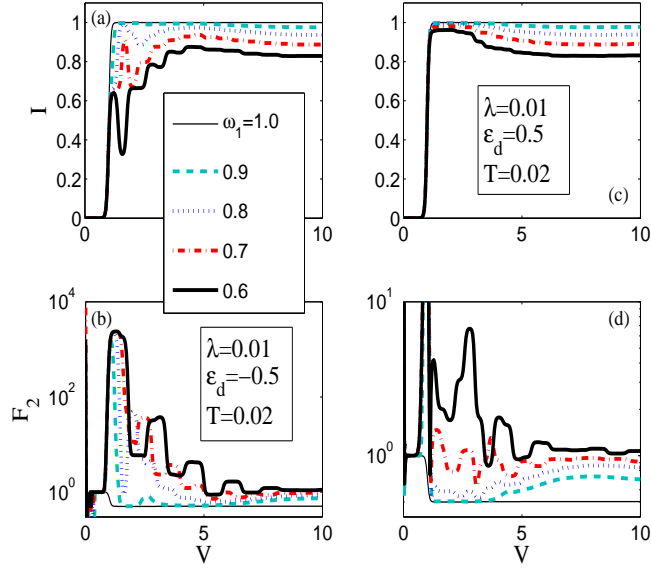}
\caption{(Color online). Distortion effect on the bias voltage dependent current, $I$ (a,c) and Fano factor, $F_2=S_2/2I$ (b,d) 
for the case of an extremely weak shift of the equilibrium position of the vibrational potential, $\lambda=0.01$ with 
$\varepsilon_d=-0.5$ (a,b) and $\varepsilon_d=0.5$ (c,d), $\Gamma_L=\Gamma_R$ and with no environmental dissipation. The 
temperature is $T=0.02$.}
\label{figif}
\end{figure}

\begin{figure}[htb]
\includegraphics[height=7.5cm,width=8.5cm]{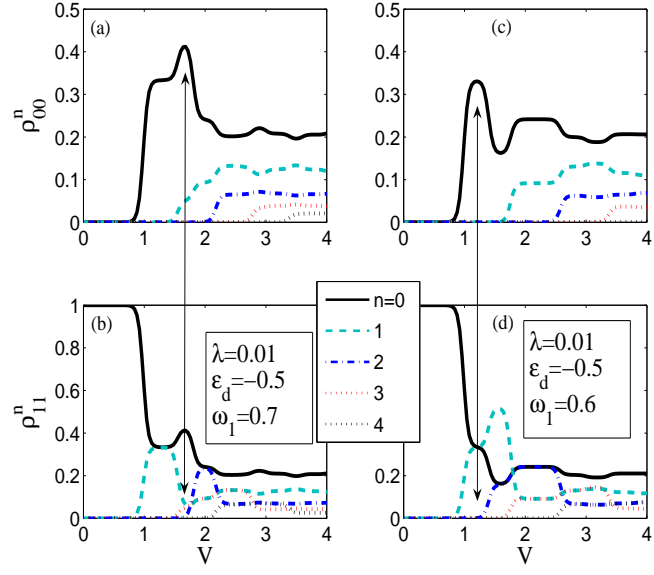}
\caption{(Color online). Joint electron-phonon occupation probabilities, $\rho_{00}^n$ (a,c) and $\rho_{11}^n$ (b,d), vs. bias 
voltage relevant to the parameters of Fig.~\ref{figif}; for (a,b) $\omega_1=0.7$ and for (c,d) $\omega_1=0.6$.}
\label{figp1}
\end{figure}

\begin{figure}[htb]
\includegraphics[height=4cm,width=8.5cm]{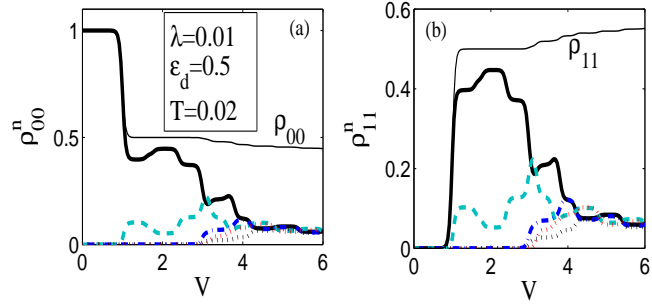}
\caption{(Color online). Joint Electron-phonon occupation probabilities, $\rho_{00}^n$ (a) and $\rho_{11}^n$ (b), vs. bias 
voltage relevant to the parameters in Fig.~\ref{figif} for $\omega_1=0.7$ and $\varepsilon_d=0.5$. The thin lines in (a) and (b) 
denote the results for $\rho_{00}$ and $\rho_{11}$, respectively.}
\label{figp2}
\end{figure}

\begin{figure}[htb]
\includegraphics[height=7.5cm,width=8.5cm]{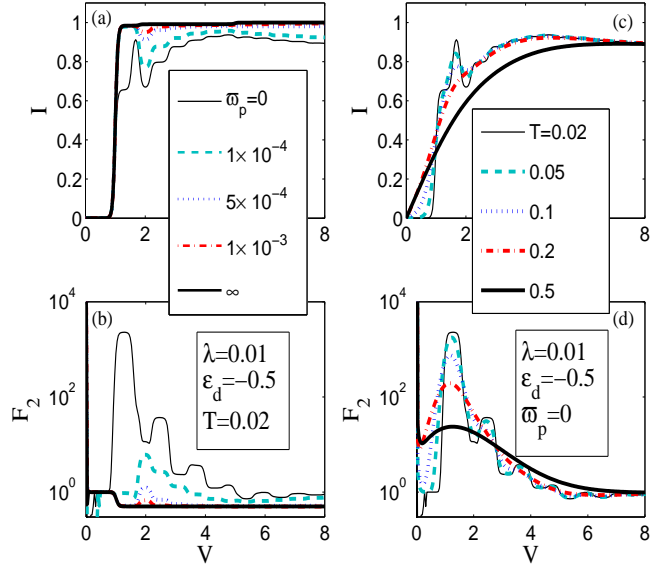}
\caption{(Color online). Calculated current $I$ (a,c), Fano factor $F_2$ (b,d), vs. bias voltage for the system with 
$\varepsilon_d=-0.5$, $\lambda=0.01$, and $\omega_1=0.7$. (a,b) are for the results with various environmental dissipation rates 
$\varpi_p$ at the temperature $T=0.02$; (c,d) are with different temperatures with no dissipation.}
\label{figif2}
\end{figure}

\begin{figure}[htb]
\includegraphics[height=7.5cm,width=8.5cm]{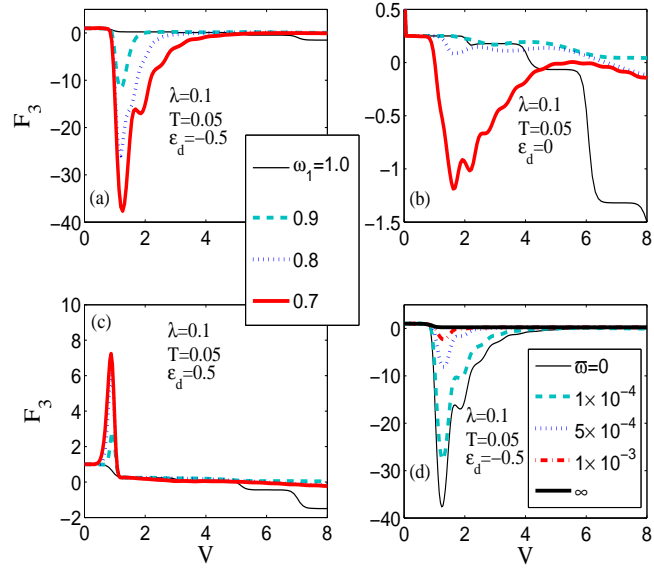}
\caption{(Color online). Distortion effect on bias-voltage-dependent skewness, $F_3=S_3/2I$, for the case of an extremely weak 
shift of the equilibrium position of the vibrational potential, $\lambda=0.1$ with $\varepsilon_d=-0.5$ (a), $0$ (b), and $0.5$ 
(c), and with no environmental dissipation. (d) Skewness vs bias voltage for $\varepsilon_d=-0.5$ and $\omega_1=0.7$ with various 
dissipation rates $\varpi_p$. The temperature is $T=0.05$.}
\label{figf31}
\end{figure}

\begin{figure}[htb]
\includegraphics[height=7.5cm,width=8.5cm]{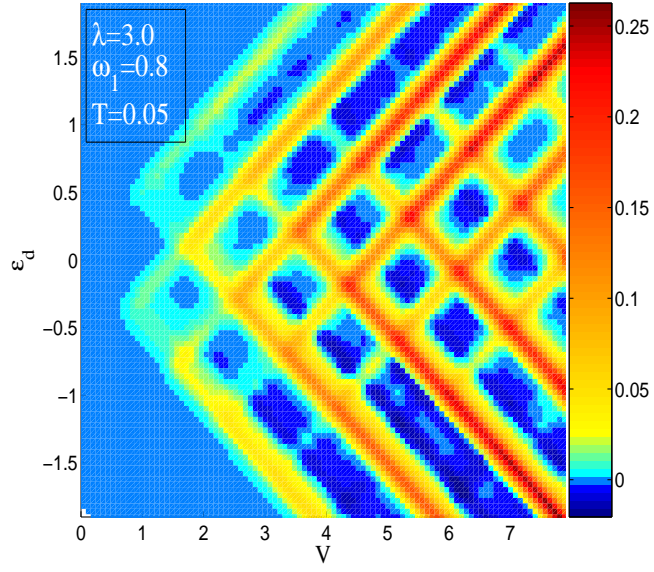}
\caption{(Color online). Differential conductance in the $V$-$\varepsilon_d$ plane for $\lambda=3.0$ and $\omega_1=0.8$ with 
$\Gamma_L=\Gamma_R$ and with no environmental dissipation. The temperature is $T=0.05$.}
\label{figf3D2}
\end{figure}

\begin{figure}[htb]
\includegraphics[height=7.5cm,width=8.5cm]{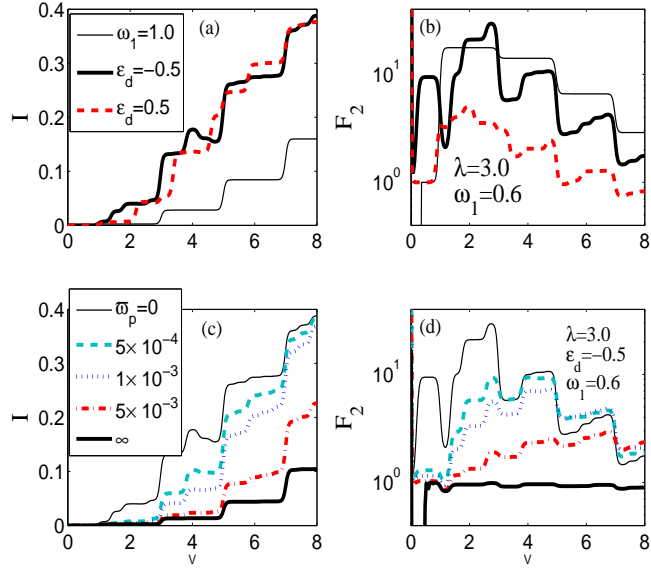}
\caption{(Color online). Current, $I$ (a,c) and Fano factor, $F_2=S_2/2I$ (b,d) vs bias voltage for the case with $\lambda=3.0$ 
and $\omega_1=0.6$. (a,b) are plotted for the cases with $\varepsilon_d=-0.5$ (solid lines), $0.5$ (dashed lines), and with no 
environmental dissipation. For comparison, we also plot the results without distortion (thin lines). (c,d) exhibit the roles of 
environmental dissipation for the system with $\varepsilon_d=-0.5$. The temperature is $T=0.02$.}
\label{figif3}
\end{figure}

\begin{figure}[htb]
\includegraphics[height=7.5cm,width=8.5cm]{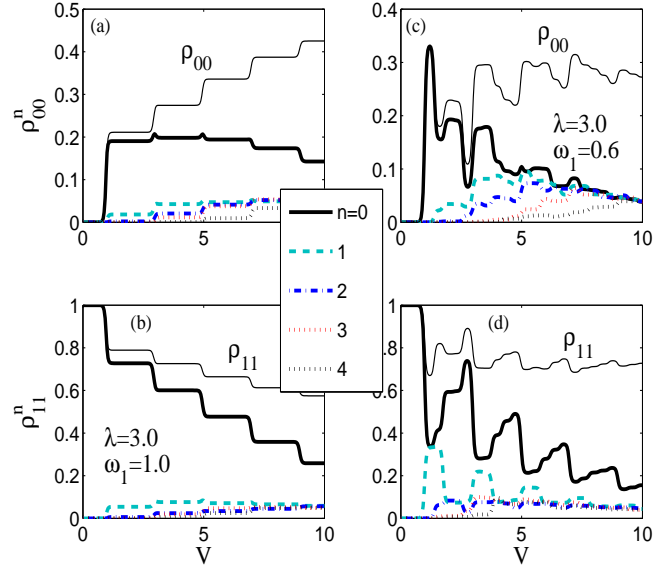}
\caption{(Color online). Joint electron-phonon occupation probabilities, $\rho_{00}^n$ (a,c) and $\rho_{11}^n$ (b,d), vs. bias 
voltage; for (a,b) $\omega_1=1.0$ and for (c,d) $\omega_1=0.6$ with $\lambda=3.0$, $\varepsilon_d=-0.5$ and $T=0.02$.}
\label{figp3}
\end{figure}

\begin{figure}[htb]
\includegraphics[height=6.5cm,width=8cm]{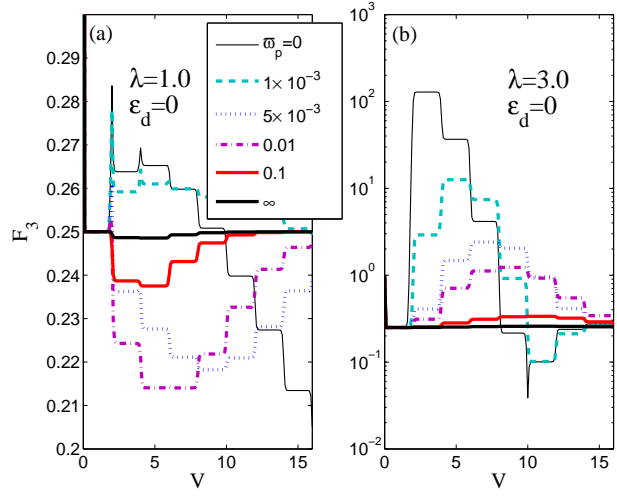}
\caption{(Color online). Bias-voltage-dependent skewness, $F_3=S_3/2I$, for the case of no distortion of the vibrational 
potential with $\varepsilon_d=0$ and $\lambda=1.0$ (a), $3.0$ (b), and with various environmental dissipation rates. The 
temperature is $T=0.02$.}
\label{figf33}
\end{figure}

\begin{figure}[htb]
\includegraphics[height=6.5cm,width=8.5cm]{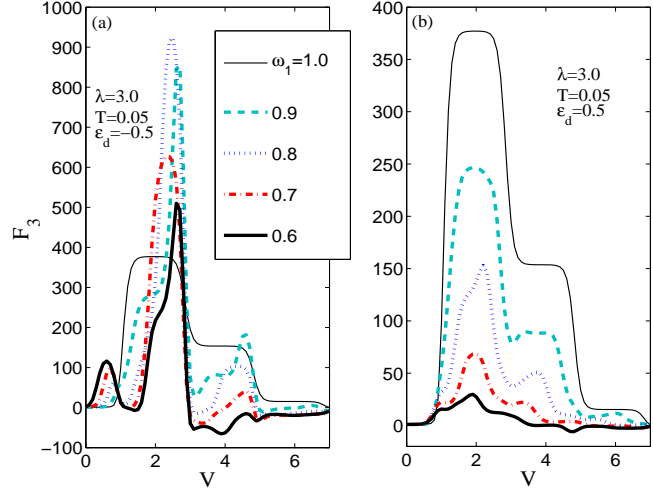}
\caption{(Color online). Distortion effect on bias-voltage-dependent skewness, $F_3=S_3/2I$, for the case of $\lambda=3.0$ with 
$\varepsilon_d=-0.5$ (a) and $0.5$ (b), and with no environmental dissipation. The temperature is $T=0.05$.}
\label{figf34}
\end{figure}

\noindent FIG.3: (Color online). Joint electron-phonon occupation probabilities, $\rho_{00}^n$ (a,c) and $\rho_{11}^n$ (b,d), vs. 
bias voltage relevant to the parameters of Fig.~\ref{figif}; for (a,b) $\omega_1=0.7$ and for (c,d) $\omega_1=0.6$.

\vspace{1cm} 

\noindent FIG.4: (Color online). Joint Electron-phonon occupation probabilities, $\rho_{00}^n$ (a) and $\rho_{11}^n$ (b), vs. 
bias voltage relevant to the parameters in Fig.~\ref{figif} for $\omega_1=0.7$ and $\varepsilon_d=0.5$. The thin lines in (a) and 
(b) denote the results for $\rho_{00}$ and $\rho_{11}$, respectively.

\vspace{1cm} 

\noindent FIG.5: (Color online). Calculated current $I$ (a,c), Fano factor $F_2$ (b,d), vs. bias voltage for the system with 
$\varepsilon_d=-0.5$, $\lambda=0.01$, and $\omega_1=0.7$. (a,b) are for the results with various environmental dissipation rates 
$\varpi_p$ at the temperature $T=0.02$; (c,d) are with different temperatures with no dissipation.

\vspace{1cm} 

\noindent FIG.6: (Color online). Distortion effect on bias-voltage-dependent skewness, $F_3=S_3/2I$, for the case of an extremely 
weak shift of the equilibrium position of the vibrational potential, $\lambda=0.1$ with $\varepsilon_d=-0.5$ (a), $0$ (b), and 
$0.5$ (c), and with no environmental dissipation. (d) Skewness vs bias voltage for $\varepsilon_d=-0.5$ and $\omega_1=0.7$ with 
various dissipation rates $\varpi_p$. The temperature is $T=0.05$.

\vspace{1cm} 

\noindent FIG.7: (Color online). Differential conductance in the $V$-$\varepsilon_d$ plane for $\lambda=3.0$ and $\omega_1=0.8$ 
with $\Gamma_L=\Gamma_R$ and with no environmental dissipation. The temperature is $T=0.05$.

\vspace{1cm} 

\noindent FIG.8: (Color online). Current, $I$ (a,c) and Fano factor, $F_2=S_2/2I$ (b,d) vs bias voltage for the case with 
$\lambda=3.0$ and $\omega_1=0.6$. (a,b) are plotted for the cases with $\varepsilon_d=-0.5$ (solid lines), $0.5$ (dashed lines), 
and with no environmental dissipation. For comparison, we also plot the results without distortion (thin lines). (c,d) exhibit 
the roles of environmental dissipation for the system with $\varepsilon_d=-0.5$. The temperature is $T=0.02$.

\vspace{1cm} 

\noindent FIG.9: (Color online). Joint electron-phonon occupation probabilities, $\rho_{00}^n$ (a,c) and $\rho_{11}^n$ (b,d), vs. 
bias voltage; for (a,b) $\omega_1=1.0$ and for (c,d) $\omega_1=0.6$ with $\lambda=3.0$, $\varepsilon_d=-0.5$ and $T=0.02$.

\vspace{1cm} 

\noindent FIG.10: (Color online). Bias-voltage-dependent skewness, $F_3=S_3/2I$, for the case of no distortion of the vibrational 
potential with $\varepsilon_d=0$ and $\lambda=1.0$ (a), $3.0$ (b), and with various environmental dissipation rates. The 
temperature is $T=0.02$.

\vspace{1cm} 

\noindent FIG.11: (Color online). Distortion effect on bias-voltage-dependent skewness, $F_3=S_3/2I$, for the case of 
$\lambda=3.0$ with $\varepsilon_d=-0.5$ (a) and $0.5$ (b), and with no environmental dissipation. The temperature is $T=0.05$.

\newpage

\newpage

\newpage

\newpage

\newpage

\newpage

\newpage

\newpage

\newpage

\newpage

\newpage

\end{document}